\documentclass{ws-procs9x6}

\begin{document}

\title{TWO-PHOTON EXCHANGE IN ELECTRON-PROTON ELASTIC SCATTERING: THEORY UPDATE}

\author{Andrei V.~AFANASEV}

\address{Department of Physics, Hampton University, Hampton, VA 23668, USA\\
and\\
Theory Center, Jefferson Lab, Newport News, VA 23606, USA}

\begin{abstract}
Recent theoretical developments in the studies of two-photon exchange effects in elastic electron-proton scattering are reviewed. 
Two-photon exchange mechanism is considered a likely source of discrepancy between polarized and unpolarized experimental 
measurements of the proton electric form factor at momentum transfers of several GeV$^2$. This mechanism predicts measurable effects 
that are currently studied experimentally. 
\end{abstract}

\vspace{1cm} 

\bodymatter

Measurements of elastic nucleon form factors reached a new level of accuracy, with separation of electric and magnetic contributions made possible
at high transferred momenta. At Jefferson Lab, due to a 100\% duty factor of the electron beam and implementation of nucleon spin polarization techniques,
electric nucleon form factors were measured up to 4-momentum transfers $Q^2$= 5.6 GeV$^2$ for the proton \cite{exp_ff_proton} and $Q^2$= 1.45 GeV$^2$ for the neutron 
\cite{exp_ff_neutron}. Extension of the measurements up to  $Q^2$= 9 GeV$^2$ via recoil proton polarimetry is underway \cite{recoil_9gev2}.

Polarization-based results, however, appeared to be in conflict with earlier 
unpolarized cross section measurements at SLAC \cite{SLACexp}.  In high-$Q^2$ kinematics, the difference between
the measured values of the proton electric/magetic form factor ratio, $G_{Ep}/G_{Mp}$, was as large as a factor of five, 
resulting in important theoretical and phenomenological implications, 
{\it c.f.} Ref.~\cite{HydeWright:2004gh}. The observed discrepancy between unpolarized and polarized experimental techniques prompted 
new cross section measurements at Jefferson Lab \cite{NewRos}. These later measurements also appeared to 
be in conflict with polarization data, confirming a systematic difference between the data from the different experimental
techniques.

The resolution of this conflict was suggested \cite{GVDH} to be due to a higher-order electromagnetic effect of two-photon exchange
not accounted for in the experimental analysis. Model calculations \cite{Blunden1} and \cite{Chen} lead to similar conclusions, attributing
over a half of experimental discrepancy to two-photon exchange corrections. A detailed account of the status of theory and experiment
in two-photon exchange can be found in the recent reviews, Refs.\cite{CVdH,Arr07}. 
Here, I present an update on the status of the two-photon exchange problem and its implications, and highlight related theoretical issues.

%
%
\begin{figure}[t]
\begin{center}
\includegraphics[width=0.7\textwidth]{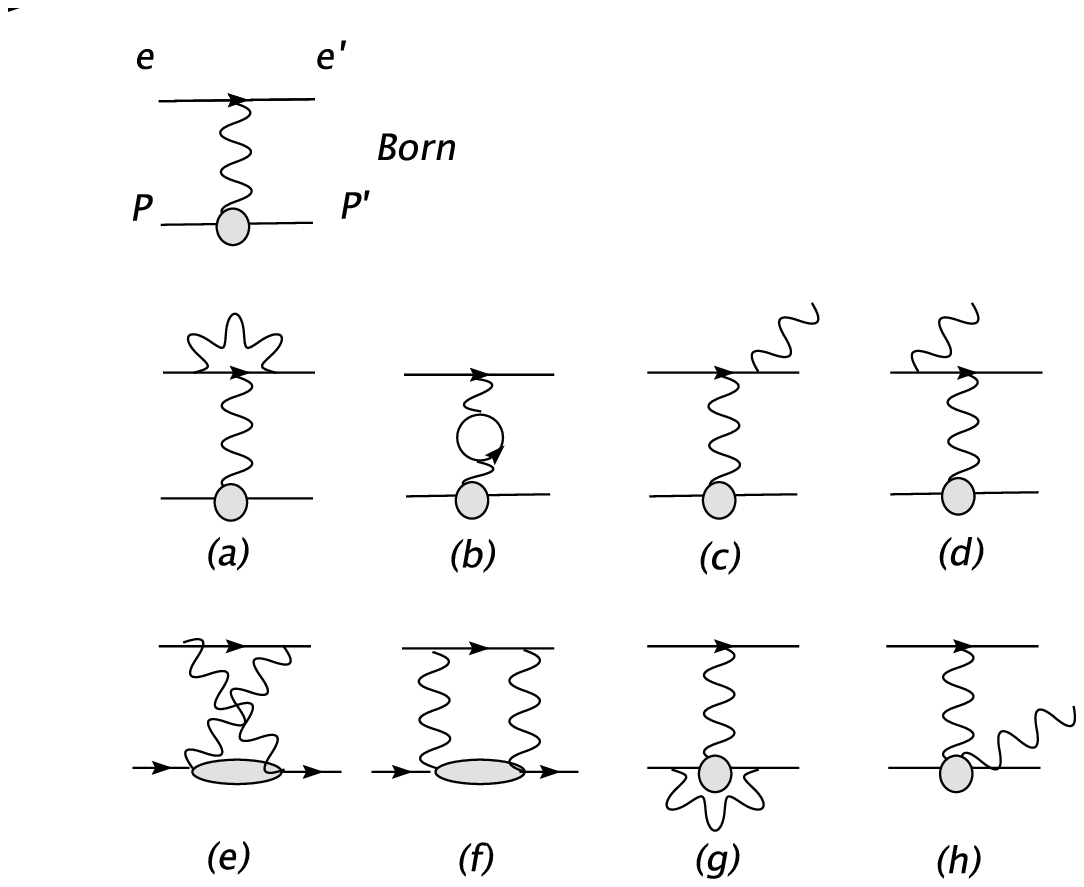}
\end{center}
\caption[]{QED radiative corrections for electron-proton scattering:
(a)~Electron vertex correction; (b)~Vacuum polarization; (c,d) Electron bremsstrahlung; (e,f) Two-photon exchange; (g) Proton vertex correction;
and (f) Proton bremsstahlung (virtual Compton scattering).}
\label{RCdiagrams}
\end{figure}
Let us start with reviewing a full set of higher-order electromagnetic corrections to electron-proton scattering Fig.\ref{RCdiagrams}.
Contributions of the diagrams Fig.\ref{RCdiagrams}a,c,d can be calculated using standard QED techniques. They are enhanced by
large logarithmic factors $~log \frac{Q^2}{m_e^2}$, resulting in radiative corrections of the order of tens per cent that in addition
depend on details of experimental cuts in the phase space of the radiated photon and electron scattering angles (at fixed $Q^2$). 
Vacuum polarization, Fig.\ref{RCdiagrams}b,
albeit has model uncertainties due to hadronic loop contributions, does not alter angular
dependence of cross section at {\it fixed} momentum transfer $Q^2$, and hence it has no impact on Rosenbluth separation.
Subprocesses with an additional photon coupling only to the proton Fig.\ref{RCdiagrams}g,f show negligible angular dependence when constrained by 
kinematic cuts of elastic scattering. 

The bremsstrahlung correction of  Fig.\ref{RCdiagrams}c,d was calculated in Ref.\cite{Tsai61} in soft-photon approximation,
and this result was applied in data analysis in Ref.\cite{SLACexp}. If this contribution is calculated fully
including also hard-photon emission, for example,  according to Ref.\cite{AAM01} or Ref.\cite{AAM00}, 
it leads to about 1 per cent additional absolute correction \cite{Afanasev_talk05} to the experimental \cite{SLACexp} 
Rosenbluth slope at Q$^2$=6 GeV$^2$. This additional correction accounts for about one fifth of the discrepancy between Rosenbluth \cite{SLACexp}
and polarization data \cite{Afanasev_talk05} when missing mass cuts on the radiated photon are chosen to match
experimental ones. The choice of kinematic cuts is essential since the magnitude of bremsstrahlung correction strongly depends on them.
For example, if one uses a generic energy cut parameter for all electron scattering angles ({\it e.g.}, c=0.97 as in Ref.\cite{BKT07}) the extracted Rosenbluth
slope reduces by about 5\% at Q$^2$=6 GeV$^2$, thereby seemingly `resolving' disagreement between Rosenbluth and polarization data.
It is therefore very important that all refined calculations of bremsstrahlung corrections are also as accurate in the choice
of kinematic cuts when compared with specific experimental analysis.

Let us take a closer look at the two-photon exchange process, Fig.\ref{RCdiagrams}e,f. In the approach developed by Tsai \cite{Tsai61},
these contributions were calculated in a limit when one of the exchanged photons carries a negligible 4-momentum. This contribution
to the scattering amplitude is (infra-red) divergent, and the divergence is cancelled at the cross-section level by adding interference 
between the bremsstrahhlung diagrams Fig.\ref{RCdiagrams}c,d and f. It therefore also depends on the details of experimental cuts on
the radiated photon kinematics. The good news is that the calculation with soft second photon exchange does not require additional knowledge on the
nucleon structure: It can be done in terms of one-photon exchange contribution times a `soft' factor that is independent on nucleon structure \cite{Tsai61}.

\begin{figure}[t]
\begin{center}
\includegraphics[width=0.6\textwidth]{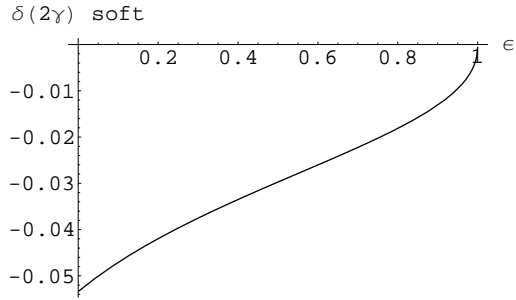}
\end{center}
\caption{`Soft' two-photon exchange correction combined with interference of electron and proton bremsstrahlung according to Ref. \cite{Tsai61}.
The relative radiative correction to the cross section is plotted aganist the standard kinematic variable $\epsilon$ for $Q^2=$ 6 GeV$^2$.}

\label{soft2g}
\end{figure}

As opposed to bremsstrahlung and vertex corrections, two-photon exchange is not enhanced by large logarithms. It is instructive to see the effect of the soft-photon-exchange portion
on the Rosenbluth plot. In Fig.\ref{soft2g}, its effect on the cross section is shown for the kinematics of SLAC experiment \cite{SLACexp} at
$Q^2=$ 6 GeV$^2$. The correction  is angular-dependent, varying between about -5\% for backward scattering and 0 for forward scattering
angles. The Rosenbluth slope measured at SLAC \cite{SLACexp} at $Q^2=$ 6 GeV$^2$ was close to 5\% with the above correction $included$.
It emphasizes importance of two-photon exchange:  Without this correction included, the extracted value of electric proton form 
factor  \cite{SLACexp} would be about a factor of $\sqrt{2}$ larger!
 
%
%

Let us estimate the effect of terms neglected in the two-photon exchange in Ref.\cite{Tsai61}. Using a well-established QED results for a structureless
spin-1/2 (muon) target \cite{Gorshkov67}, 
we find the correction for backward-angle electron scattering on a quark with a mass $m_q$ and a charge $e_q$:
\begin{equation}
\delta^{2\gamma}=-\frac{e_q \alpha}{2 \pi} \log^2\frac{s}{m_q^2},
\end{equation}
where $s$ is a Mandelstam variable, and $\alpha$ is a fine structure constant. The corresponding correction is zero
for forward electron scattering. Note that the correction is negative for positive-charge quarks, and it grows logarithmically with beam energy;
numerically, it is a few per cent for relevant kinematics, if a constituent quark mass of $m_q=$ 300 MeV is taken for the estimate.
Therefore this correction has the proper sign, magnitude and angular dependence to mimic a contribution of electric form factor to
the cross section of electron-proton scattering. 

\begin{figure}[t]
\begin{center}
\includegraphics[width=0.8\textwidth]{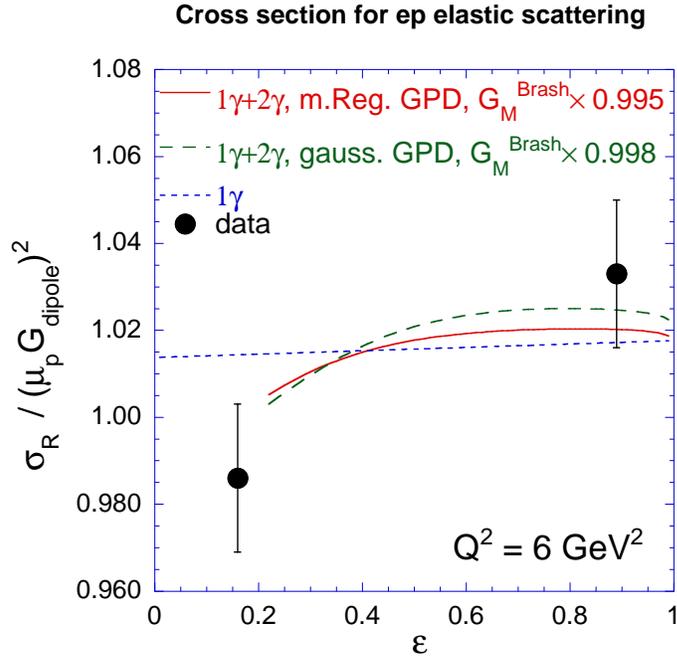}
\end{center}
\caption{Reduced ep-scattering cross section at $Q^2=$ 6 GeV$^2$. Data points are from Ref.\cite{SLACexp}. The dotted line shows an expected
result from one-photon exchange using $G_{Ep}$ fit to polarization data \cite{exp_ff_proton}; solid and dashed curves have the two-photon
exchange mechanism included within partonic approach \cite{Chen,ABC} using different models of GPD. }

\label{Ros_GPD}
\end{figure}

It motivates more detailed studies of two-photon exchange, especially at a partonic level. Such a partonic approach was developed in Refs.\cite{Chen, ABC},
where two-photon exchange amplitude was factorized into a hard subprocess of electron-quark scattering and a soft subprocess described by
generalized parton distributions (GPD). A representative result is shown in Fig.\ref{Ros_GPD}, where it can be seen that linear $\epsilon$-dependence
of the cross section is modified by a non-linear contribution from two-photon exchange. A dotted line labeled `1$\gamma$'  is an expectation
from a pure one-photon exchange mechanism with a proton electric form factor taken from polarization measurements \cite{exp_ff_proton}.
Noticing that at the same time two-photon exchange does not significantly alter interpretation of polarization data, we conclude that
within the considered model, this mechanism partially reconciles results of experimental techniques using polarized and unpolarized scattering.

In a different approach \cite{Blunden1,Blunden2}, the virtual Compton amplitude entering
the two-photon exchange mechanism was approximated with nucleon pole diagrams with on-shell form factors substituted in photon-nucleon vertices.
Despite of different dynamical models for the nucleon Compton amplitudes,
the conclusions of Refs.\cite{Blunden1,Blunden2} and Refs.\cite{Chen,ABC} are in qualitative agreement. Addition of
$\Delta$-excitation mechanism \cite{Blunden_delta} to the approach of Refs.\cite{Blunden1,Blunden2} somewhat reduced
the predicted magnitude of the two-photon effect. Higher nucleon resonances are estimated Ref.\cite{Kondratyuk07} to contribute about an order of magnitude less
 than nucleon and $\Delta$.

Clearly, the problem of two-photon exchange, especially the real part of the amplitude, is challenging because 4-dimensional momentum integration 
in the box (and cross-box) diagrams Fig.\ref{RCdiagrams}e,f requires knowledge of the nucleon Compton amplitude over a 
broad (infinite, to be exact!) range of kinematic variables not available from experiment. 
On the other hand, theoretical models applied so far are valid only within certain kinematic regions. 
One may also try a dispersive approach that takes advantage of analyticity and unitarity of the two-photon amplitude.
In the kinematics of forward electron scattering, it is possible to reduce model uncertainties by using inelastic electroproduction
structure functions measured experimentally \cite{Misha_disp}. 
A different category of papers attempt to evaluate the two-photon exchange effect using the experimentally observed difference between
Rosenbluth and polarization data, \cite{GVDH,Arrington2004,Jain2006,Belushkin07}.

The two-photon exchange mechanism also contributes to parity-violation studies of electron scattering through interference with $Z$-boson exchange,
as was pointed out in Ref.\cite{AC05}. The effect was evaluated in Ref.\cite{AC05} in GPD framework at about 2\% for backward angles and large $Q^2$.
For smaller momentum transfers the two-photon  effect is less significant, but $Q^2$ dependence of a $\gamma Z$ box contribution was found to be essential 
\cite{Zhou:2007hr} for extraction of strange-quark effects in the proton neutral weak current. 
Authors of Ref.\cite{Zhou:2007hr} used a hadronic model \cite{Blunden1,Blunden2} and found 
that a combined effect of $2\gamma$ and $\gamma Z$ exchange on the values of $G^s_E+\beta G^s_M$ 
extracted in recent experiments can be as large as -40\% in certain kinematics. A similar calculation was also presented in Ref.\cite{Tjon:2007wx}.

A comprehensive series of experiments are either underway or in preparation at Jefferson Lab with a purpose to study two-photon exchange effects
in electron-proton scattering. Since two-photon exchange correction to electron scattering observables is proportional to an odd power of the electron charge,
it can be measured directly by comparing electron and positron scattering. This method will be used in JLab experiment E-07-005,
with a tertiary beam obtained from photoproduction of electron-positron pairs. Another JLab experiment, E-05-017, analyzes non-linearity 
of Rosenbluth plot caused two-photon exchange. Angular dependence of double-spin observables is also affected by two-photon exchange
at a few per cent level \cite{ABC}, and it is being looked for in polarization-transfer measurements (JLab experiment E-04-019).
A single-spin target asymmetry caused by two-photon exchange will be studied in JLab experiment E-05-015 on a polarized $^3$He target.

\begin{figure}[t]
\begin{center}
\includegraphics[width=0.8\textwidth]{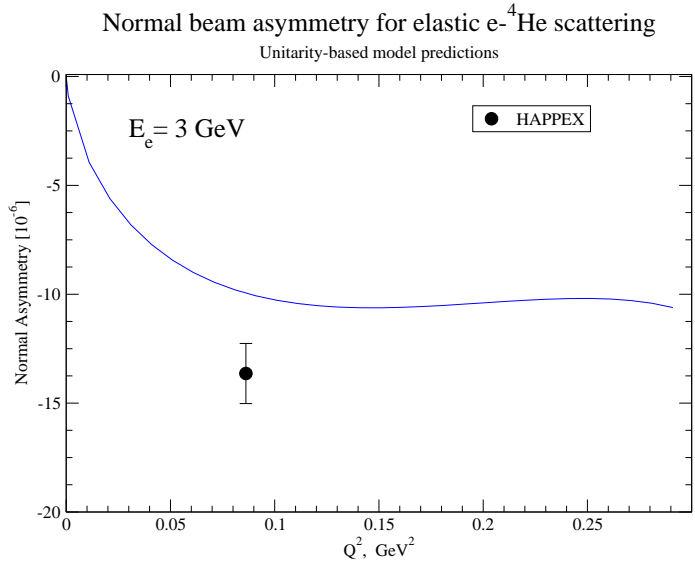}
\end{center}
\caption{Single-spin normal beam asymmetry on a $^4$He target in units of parts per million. The curve is a prediction of a unitarity-based
model \cite{AM04} extended to a nuclear target, with total photoproduction cross section and Compton $t$-slope on $^4$He used for input.
 Experimental data point is from Ref.\cite{Kaufman:2007}. Contribution of Coulomb
distortion is below a few parts per {\it billion} in the shown kinematics.}

\label{Bssa_He4}
\end{figure}

So far, the only definitive experimental observation of two-photon exchange effects came from the measurements of normal beam asymmetry at MIT-Bates \cite{Wells:2000},
MAMI \cite{Maas:2005}, and JLab \cite{Armstrong:2007,Kaufman:2007}. The observations appear to be in reasonable agreement 
with theoretical calculations at lower energies \cite{AAM02,Pasquini}, nucleon resonance region \cite{Pasquini} and above the resonance region 
\cite{AM04,Misha_disp,Misha_bssa}. Note that a unitarity-based approach \cite{AM04} applied for a nuclear $^4$He target agrees both in sign and magnitude
with recent measuments from JLab HAPPEX collaboration \cite{Kaufman:2007}, see Fig.\ref{Bssa_He4}; while the prediction in the kinematics of upcoming
PREX measurement (JLAB E06-002) on Pb-target is about -5 ppm. At the same time, this asymmetry appears to be several orders
of magnitude larger than predictions from a known mechanism of Coulomb distortion for small-angle electron scattering kinematics \cite{Cooper:2005}. 

We conclude that the two-photon exchange effect stands as a possible source of the difference between Rosenbluth and polarization techniques
for proton electric form factor measurements. It has to be included in the analysis of other precision experiments.
The two-photon exchange mechanism leads to new effects that can be studied experimentally.

Notice: Authored by Jefferson Science Associates, LLC under 
U.S.\ DOE Contract No.~DE-AC05-06OR23177. The U.S.\ Government retains a
non-exclusive, paid-up, irrevocable, world-wide license to publish or
reproduce this manuscript for U.S.\ Government purposes.
\bibliographystyle{ws-procs9x6}


\end{document}